\documentclass[11
pt]{article}
\usepackage[utf8]{inputenc}
\usepackage{lmodern}
\usepackage{amsmath}
\usepackage{fixmath}
\usepackage{exscale}
\usepackage{accents}
\usepackage{bigints}
\usepackage{physics}
\usepackage{commath}
\usepackage{float}
\usepackage{soul}
\usepackage{color}
\usepackage{subfigure}
\usepackage{caption}
\usepackage{graphicx}
\usepackage{upgreek}
\usepackage{subfigure}
\usepackage{textcomp}
\usepackage{amssymb}
\usepackage{authblk}
\usepackage[title]{appendix}
\usepackage[a4paper, total={160mm,240mm}]{geometry}
\usepackage[authoryear,round]{natbib}
\usepackage{bm}
\usepackage{authblk}

\usepackage{moreverb}

\usepackage[colorlinks=true, urlcolor=blue, pdfborder={0 0 0},citecolor=blue]{hyperref}

\newcommand\BibTeX{{\rmfamily B\kern-.05em \textsc{i\kern-.025em b}\kern-.08em
T\kern-.1667em\lower.7ex\hbox{E}\kern-.125emX}}

\title{Inverse Identification of Surface Elastic Parameters in Soft Solids Using GA-ANN Surrogate Model}
\author[1]{Md. Ayaz}
\author[2]{Abhishek Ghosh}
\author[3]{Andrew McBride}
\author[4]{M. Rashid Zafar\thanks{Corresponding author: rashidzafar@zhcet.ac.in}}
\author[4]{Md. Irfan Ansari}

\affil[1]{Civil Engineering Section, University Polytechnic, AMU, Aligarh, UP-202002, India.}
\affil[2]{ Zienkiewicz Institute for Modelling. Data and AI, Swansea University, Bay Campus, Swansea, United Kingdom}
\affil[3]{Glasgow Computational Engineering Centre, James Watt School of Engineering, University of Glasgow, United Kingdom}
\affil[4]{Department of Mechanical Engineering, ZHCET, AMU, Aligarh, UP-202002, India.}
\date{}
\begin{document}
\maketitle

\maketitle
\begin{abstract}

Surface elasticity plays a crucial role in the mechanics of soft solids at the scale of micrometers and may become significant even at the scale of millimeters in exceptional cases. However, despite the efforts made for the identification of surface elastic parameters, accurately determining these values remains challenging due to the complex interplay with bulk elasticity and nonlinearity.
To address this issue, a novel GA-based optimization framework is developed by employing an ANN-based surrogate forward model for efficient identification of surface parameters from the force–displacement response of a cylindrical specimen.
The ANN is trained on force-displacement data of soft cylindrical specimen, generated by nonlinear FE model predictions incorporating model elastic surfaces. 
The data used for training is generated for non-dimensional values of surface tension in the range of $0-5$ and surface shear modulus in the range of $0-50$. 
The accuracy and reliability of the trained ANN are established. The key novelty lies in replacing analytical forward models, which rely on idealized boundary conditions, with a data-driven surrogate trained on FE simulations under realistic constraints.
The parameter identification is performed for a number of surface parameter sets using numerically-generated force-displacement data as well as for noisy data obtained by adding $5\%$ error in simulated data.
The maximum error in identified values of parameters in all the cases is less than $8\%$. Repeatability and uncertainty analyses based on multiple noisy realizations further demonstrate the robustness of the approach, yielding narrow confidence intervals for the predicted parameters. 
The proposed framework provides a novel, efficient, and robust alternative to FE-based inverse identification of surface parameters, particularly for experimentally relevant boundary conditions.

\end{abstract}

\section{Introduction}
Soft solids with elastic modulus in the range of sub-kPa to few hundreds of kPa play a major role in emerging technologies such as soft robotics \citep{Lopez-Diaz2024}, flexible electronics \citep{Wang2018}, adhesives \citep{creton2003}, and soft tissue implants \citep{xu2023}. While the mechanical behavior of conventional solids are only significantly affected by surface stresses at the nanoscale \citep{sharma2004, miller2000}, soft solids exhibit pronounced surface stress effects even at length scales extending to several hundred microns \citep[see, {\it e.g.},][]{mora2010, style2013, mosli2025}. Whereas liquids exhibit deformation-independent isotropic surface stresses (also referred to as surface tension), surface stresses in solids are deformation-dependent \citep{jensen2017}. For example, \cite{Bain2021} have investigated the effect of stretching on the topography of silicon gels and found that the surface stress changes with deformation.

A rigorous mathematical framework for elastic surfaces was developed by \cite{gurtin1975} and extended by \cite{steigmann1999}, \cite{huang2006}, and \cite{sharma2011}, among others. In parallel with these theoretical developments, numerical modeling strategies for elastic surfaces have also been proposed. For example, \cite{parks2006}, \cite{she2009} and \cite{tian2007a} approximately solved the relations of surface elasticity using the finite element (FE) method for the infinitesimal deformation regime. Steinmann and co-workers have systematically incorporated various hyperelastic surface material models into FE formulations \citep{javili2009,javili2010,mcbride2015}. Related comparisons between atomistic and surface-enhanced continuum approaches have also highlighted the effectiveness of continuum surface models in capturing surface effects with significantly lower computational cost \citep{davydov2013}.

A number of studies in the literature have focused on quantifying surface stresses and surface elastic parameters. For example, \cite{nadermann2013} used a pendant droplet under a thin film to measure the surface tension of a solid while \cite{mondal2015} estimated the solid-liquid interfacial tension using a cylindrical microchannel in a film. \cite{xu2017} subsequently identified both the surface tension and surface elastic parameter of a silicon gel using a soft adhesive contact experiment. \cite{kim2021} have proposed a method to evaluate the surface tension of solids in the presence of hysteresis in contact angle of droplets. \cite{heyden2022} developed several mechanical tests to quantify surface elasticity in the small deformation regime.

Soft solids in most practical applications undergo large deformations, necessitating a nonlinear description of their mechanical response. This, in turn, calls for the calibration of nonlinear elastic parameters of soft solids and their surfaces.  \cite{zafar2021a} have derived analytical expressions for force and torque corresponding to various levels of stretch and twist for cylindrical specimens incorporating hyperelasticity of both bulk and surface. They have considered ideal boundary conditions at the ends (i.e., the end cross-sections are free to slip in the radial direction). However, experimentally, the cylinder ends are constrained to deform in radial direction, which limits the validity of the derived analytical expressions over length-to-radius ratio.

The use of machine learning (ML) has been increasingly applied in elasticity. For instance, ML has been used to get information about mathematical expressions for hyperelastic energy functional from sparse experimental data \citep{bahmani2024}.
Prediction of mechanical properties in metamaterials and the inverse problem of determining elastic modulus distributions are another couple of examples \citep{guo2021,zhou2022}. Parameter identification in solid mechanics falls under the category of inverse problems that are formulated as an optimization problem and the solution procedure involves solution of forward problem several times \citep{haghighat2021}. 
Typically, it is not possible to obtain a closed form solution of forward problems and one need to approach numerical methods. For example, \cite{cooreman2007} discussed about FE based inverse method in order to obtained elasto-plastic material parameters. Furthermore, \cite{hwang2009} combined the merits of genetic algorithm (GA) and simulated annealing optimization to identify the elastic constants of composite plates using FE simulation to solve the forward problem. The computational cost associated with repeatedly solving such forward problems motivates the development of efficient surrogate models, such as those based on ML approaches.

In this study, we trained an Artificial Neural Network (ANN) on force–displacement data for a cylindrical specimen generated using the nonlinear FE framework of \cite{zafar2021}, covering a broad range of surface and bulk elastic parameters. The trained ANN eliminates the limitations arising from the idealized assumption of radial slip at the cylinder ends in the analytical solution of \cite{zafar2021a}, as it is trained on FE data generated under realistic boundary conditions with constrained radial displacement.
It provides a fast, computationally efficient forward model compared with the FE simulation. 
The key novelty of this work lies in the integration of the ANN-based surrogate model within a GA-based optimization framework for inverse identification of surface parameters from experimental or synthetic force-displacement data.

Although physical experiments are beyond the scope of the present work, controlled noise is added to the FE-simulated data to mimic experimental conditions and assess the robustness of the proposed parameter identification framework. While the calibration of bulk elastic parameters in soft solids is well established, identifying surface stiffness parameters remains challenging and is widely recognized as a difficult problem \citep{davydov2013}. The proposed framework therefore offers an efficient alternative for calibrating surface parameters.

The main contributions of this work are as follows:
\begin{enumerate}
    \item development of a data-driven ANN surrogate trained on FE simulations under realistic boundary conditions, overcoming limitations of existing analytical models,
    \item integration of the surrogate model within a GA-based inverse framework for parameter identification and
    \item systematic assessment of robustness under noisy data conditions.
\end{enumerate}

The organization of the article is as follows. Section~\ref{Sec:theory} briefly presents the theoretical framework of hyperelastic solids and surface elasticity. The generation of force–displacement data using nonlinear FE simulation is described in Section~\ref{Sec:FEM}. The results of the forward ANN predictions and the inverse identification of surface parameters are presented and discussed in Section~\ref{Sec:Results}. Finally, concluding remarks are provided in Section~\ref{Sec:conclusion}.

\section{Theory of bulk and surface elasticity}
\label{Sec:theory}
Consider a body with initial configuration $\mathcal{B}_o$ that deforms into the current configurations $\mathcal{B}$ at time $t$. The boundary of the body in the initial and current configurations is denoted by $\partial\mathbf{B}_o$ and $\partial\mathcal{B}$, respectively. A part of the boundary in the initial configuration $\mathcal{S}_o\subset\partial\mathcal{B}_o$ is having surface elasticity which maps into $\mathcal{S}\subset\partial\mathcal{B}$ in the deformed configuration. A point $P$ with position vector $\bf{X}\in\mathcal{B}_o$ maps to position $\bm{x}(\bm{X},t)\in\mathcal{B}$. The deformation gradient tensor (for bulk) is defined by
\begin{eqnarray}
    \bm{F}=\nabla\bm{x}=\dfrac{\partial\bm{x}}{\partial\bm{X}} \, ,
\end{eqnarray}
while, the surface deformation gradient is defined by
\begin{eqnarray}
    \bm{F}_s=\bm{F}\bm{I}_s,
\end{eqnarray}
with $\bm{I}_s$ being the inclusion map \citep{gurtin1975,zafar2021,ZAFAR2022} which maps any vector in the tangent space of the undeformed surface to the same vector in three-dimensional vector space. For instance, if the tangent space at particular point on the surface is spanned by orthonormal basis $\bm{g}_1$ and $\bm{g}_2$, then inclusion map is
\begin{eqnarray}
    \bm{I}_s=\begin{bmatrix}
        (\bm{g}_1\cdot\bm{e}_1) & (\bm{g}_2\cdot\bm{e}_1)\\
        (\bm{g}_1\cdot\bm{e}_2) & (\bm{g}_2\cdot\bm{e}_2)\\
        (\bm{g}_1\cdot\bm{e}_2) & (\bm{g}_2\cdot\bm{e}_3)
    \end{bmatrix},
\end{eqnarray}
where, $\bm{e}_i$ represents the orthonormal basis vectors for three-dimensional vector space. The left and right Cauchy-Green deformation tensors for bulk and surface are defined, respectively, as
\begin{eqnarray}
    \bm{C}=\bm{F}^T\bm{F},\nonumber\\
    \bm{C}_s=\bm{F}^T_s\bm{F}_s.
\end{eqnarray}
\begin{figure}
    \begin{center}
    \includegraphics[width=0.6\linewidth]{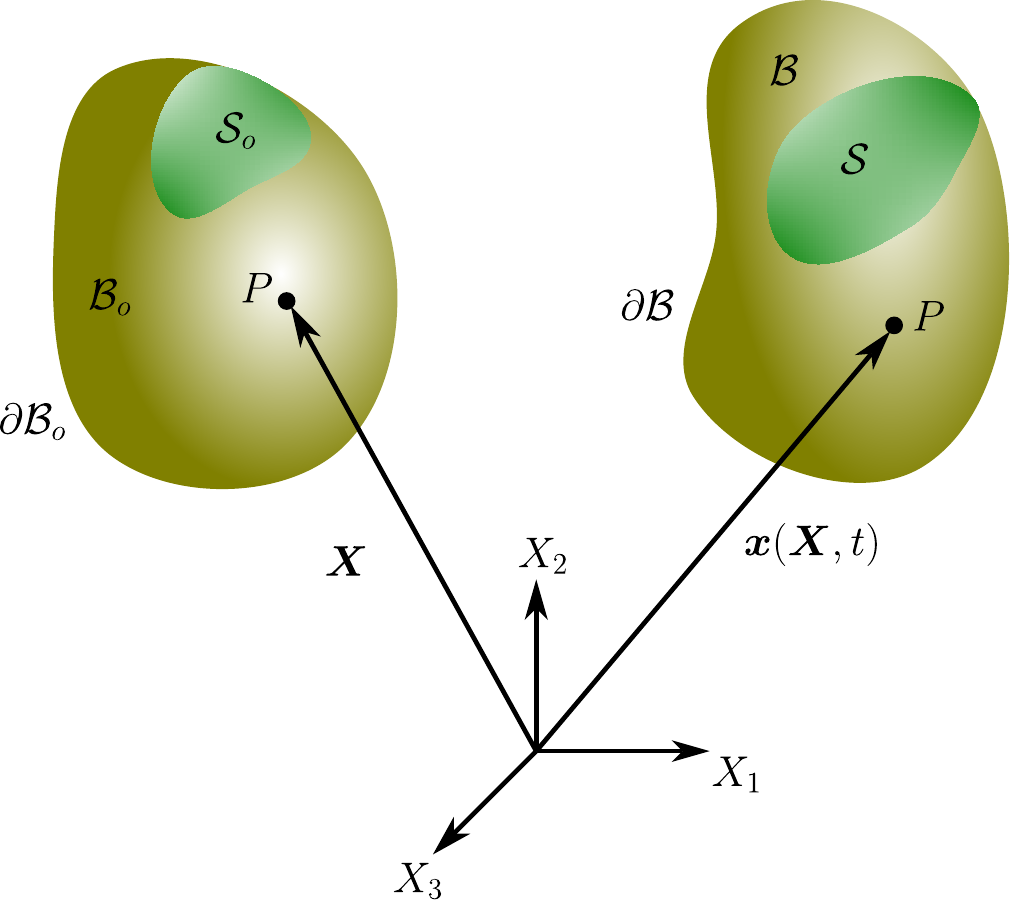}
    \caption{Undeformed $(\mathcal{B}_o)$ and deformed ($\mathcal{B}$) configurations of a general body. Surface $\mathcal{S}_o\in\partial\mathcal{B}_o$ is elastic which maps to $\mathcal{S}$ in deformed configuration.}
    \label{fig:mapping}    
    \end{center}
\end{figure}

The strain energy density functional $\psi$ for an isotropic hyperelastic bulk is dependent on the invariants of $\bm{C}$ \citep{Holzapfel2000}. Within the scope of this work, an incompressible neo-Hookean material model is considered, i.e.,
\begin{eqnarray}
    \psi=\dfrac{\mu}{2}(I_1-3),
\end{eqnarray}
with $I_1$ being the first invariant of $\bm{C}$ and $\mu$ the shear modulus. 
The surface strain energy density is modeled using a neo-Hookean-type formulation proposed by \cite{javili2010}. 
Although the original formulation includes two elastic terms, only one term is retained here, since the coefficient associated with the omitted term was shown to have a similar effect \citep{zafar2021}. Accordingly, the surface strain energy density chosen is given by
\begin{eqnarray}
    \gamma=\sigma_o\sqrt{I_2^s}+\dfrac{\mu_s}{2}(I_1^s-2-\log I_2^s).
\end{eqnarray}
Here, the first term is associated with surface energy of liquid-like surfaces, while the second term is associated with elasticity. The quantities $\sigma_o$ and $\mu_s$ are surface parameters surface tension and surface shear modulus, respectively.
$I_1^s=\textrm{tr}(\bm{C}_s)$ and $I_2^s=\textrm{det}(\bm{C}_s)$ are first and second invariants of $\bm{C}_s$. Furthermore, Cauchy stress tensors for the bulk and surface are defined as
\begin{eqnarray}
    \bm{\sigma}=-p\bm{1}+2\bm{F}\dfrac{\partial \psi}{\partial \bm{C}}\bm{F}^T,\nonumber\\
    \bm{\sigma}^s=\dfrac{2}{\sqrt{I_2^s}}\bm{F}_s\dfrac{\partial \gamma}{\partial \bm{C}_s}\bm{F}^T_s,
\end{eqnarray}
where, $p$ is Lagrange multiplier associated to the incompressibility constraint of the bulk and is identified as hydrostatic pressure.

\section{Force-Extension Data for Cylindrical Specimen}
\label{Sec:FEM}
\begin{figure}
    \begin{center}
    \includegraphics[width=0.5\linewidth]{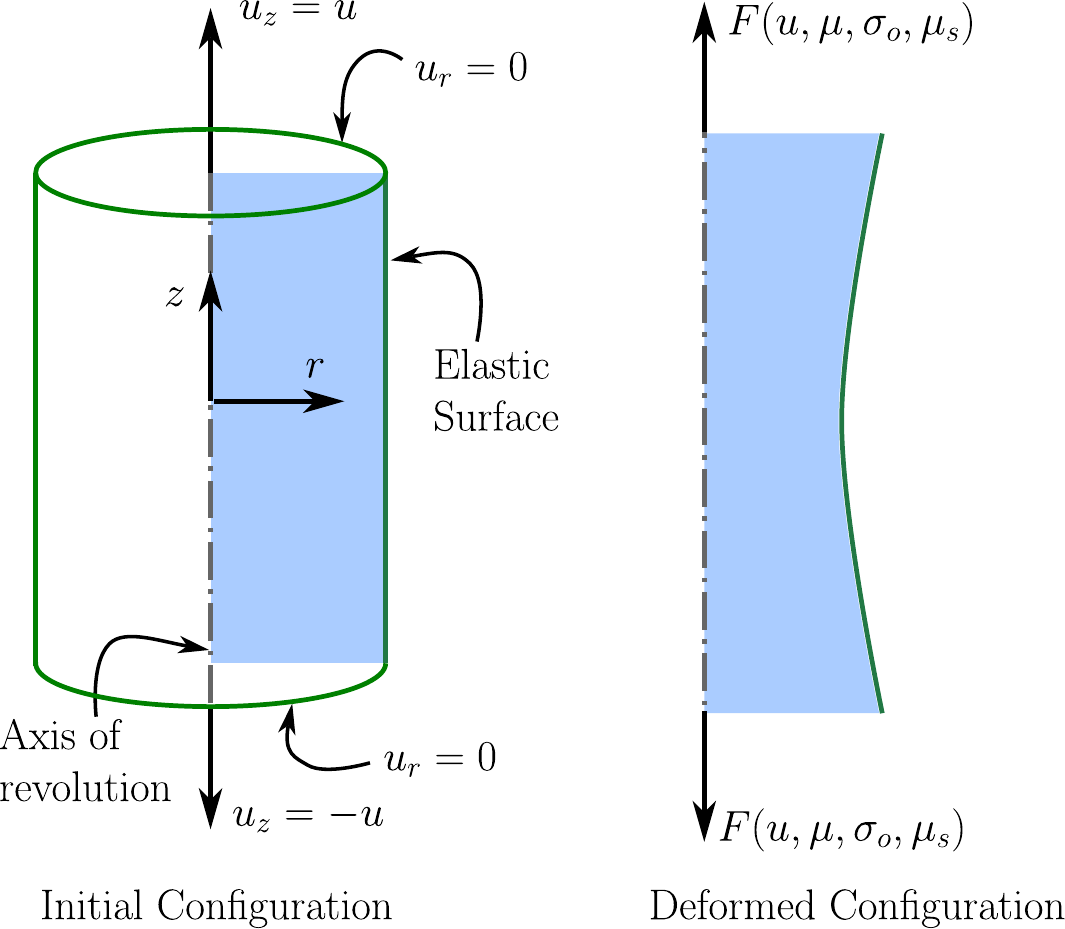}
    \caption{Axisymmetric model of a cylinder under extension with elastic surface.}
    \label{fig:axiSchematic}    
    \end{center}
\end{figure}

A nonlinear FE framework developed by \cite{zafar2021} is used for generating force-extension data for a cylindrical specimen for a range of bulk and surface parameters. An axisymmetric 2-dimensional analysis is performed in order to exploit the rotational symmetry of the problem. The schematic of boundary value problem is shown in Fig.~\ref{fig:axiSchematic}. The lower and upper ends are assigned non-zero axial displacements $u_z=-u$ and $u_z=u$, respectively, while keeping other components of displacement fixed. The bulk of the cylinder is discretized into axisymetric 8-noded quadrilateral elements with hybrid formulation to incorporate incompressibility of neo-Hookean material. In order to model surface elasticity, a 3-noded user-defined line element is implemented in the FE software Abaqus/Standard 2023 (SIMULIA, Dassault Syst\`emes, V\'elizy-Villacoublay, France). The details may be found in supplementary material provided with the paper. 

The axial force $F$ is evaluated from the simulation for each combination of displacement $u$ and material parameters $\sigma_o$ and $\mu_s$. 408 forward simulations are performed in order to generate 41208 data points.
Motivated by the study of \cite{zafar2021a}, FE simulations have been performed in a non-dimensional framework. All the lengths have been nondimensionalized by the initial radius $R$ of the cylinder, while the forces are non-dimensionalized by $\mu R^2$, i.e.,
\begin{eqnarray}
    \overline{L}&=&L/R,\nonumber\\
    \overline{u}&=&u/R,\nonumber\\
    \overline{F}&=&F/\mu R^2.
\end{eqnarray}
Similarly, the surface parameters, having dimension the same as that of $\mu R$, are non-dimensionalized as:
\begin{eqnarray}
    \overline{\sigma}_o=\sigma_o/\mu R,\nonumber\\
    \overline{\mu}_s=\mu_s/\mu R.
\end{eqnarray}
Data is generated for cylinder of length, $\overline{L}=1$ (i.e., $L=R$) and maximum displacement given to the ends are $\overline{u}=0.5$ (i.e., $u=0.5R$). The values of surface parameters $\overline{\sigma}_o$ and $\overline{\mu}_s$ for which data is generated are listed in the Table \ref{tab:surfParaRanges}.
\begin{table}[]
    \centering
    \caption{Values of Surface Parameters}
    \begin{tabular}{c|c}
    \hline
       Parameter  & Values \\
       \hline
       $\overline{\sigma}_o$  & 0, 0.01, 0.1, 1, 5 \\
       $\overline{\mu}_s$  & 0, 0.01, 0.1, 1, 10, 50\\
       \hline
    \end{tabular}
    \label{tab:surfParaRanges}
\end{table}
The maximum value of $\overline{\mu}_s$ higher than the maximum value of $\overline{\sigma}_o$ is motivated by the findings of \cite{jensen2017}.

\begin{figure}
    \begin{center}
    \includegraphics[width=0.95\linewidth]{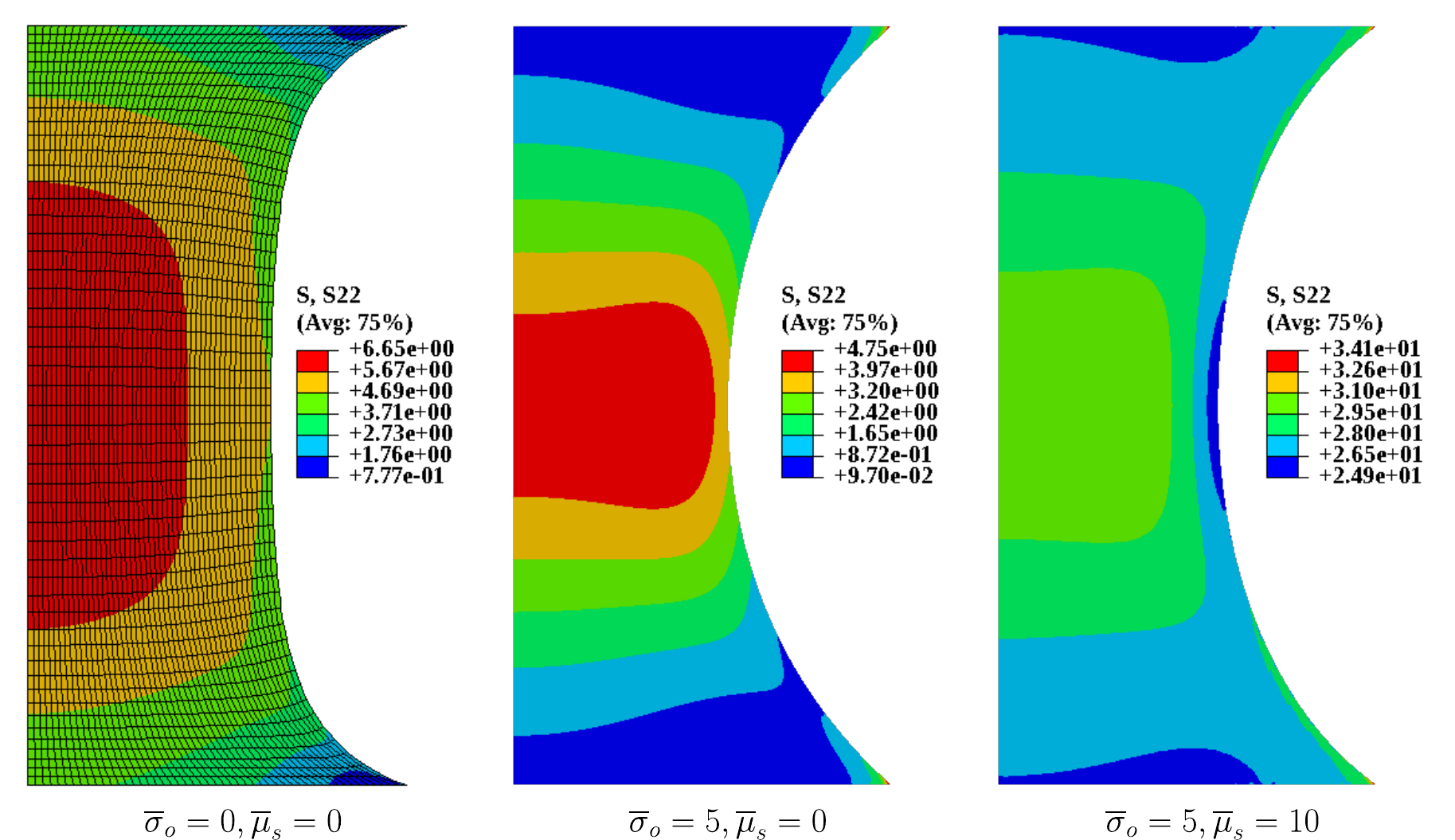}
    \caption{Deformed shapes of cylinder for three sets of surface parameters with contours representing $\sigma_{22}$ stress component. The left most subfigure shows the mesh.}
    \label{fig:deformedShapes}    
    \end{center}
\end{figure}

\begin{figure}
    \begin{center}
    \includegraphics[width=0.55\linewidth]{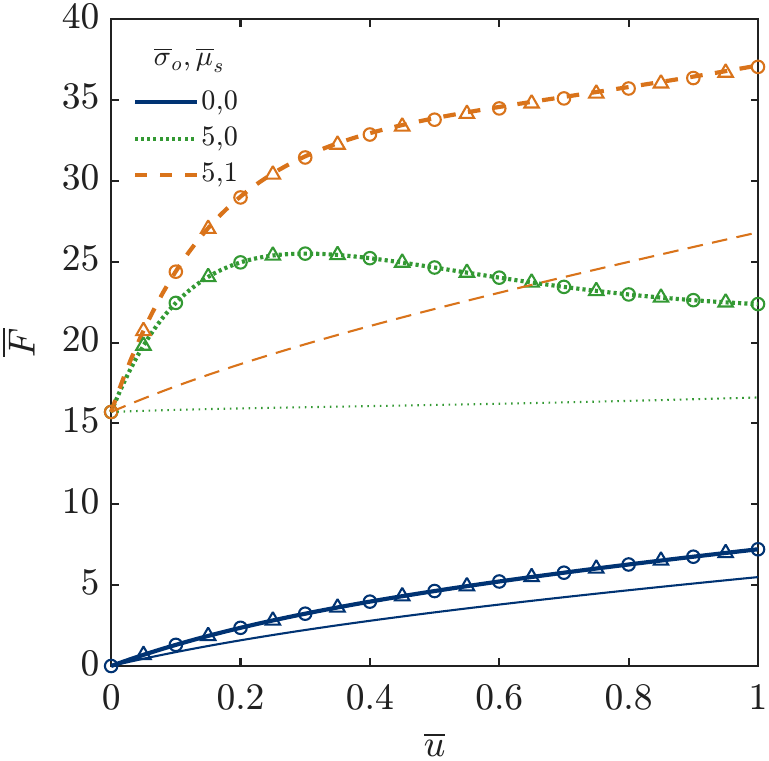}
    \caption{Force-displacement data generated through FE simulations for three sets of surface parameters. Results for three mesh refinement with element sizes $l/R=0.04,\;0.02$ and $0.01$ are represented by triangles, thick lines and circles, respectively. Thin lines are analytical results from \cite{zafar2021a} with radial slips at ends.}
    \label{fig:force-displacement-FEM}    
    \end{center}
\end{figure}

The deformed shapes of cylinder for three representative sets of surface parameters are shown in Fig.~\ref{fig:deformedShapes} for $\overline{u}=0.5$. 
It is clear from the contours of non-dimensionalized bulk Cauchy stress component (i.e., $\overline{\sigma}_{zz}=\sigma_{zz}/\mu$), that higher surface tension lowers bulk axial stress values in the stretched configuration. 
This occurs because  stretching causes axial tension while surface tension causes compressive stress. 
On the contrary to surface tension, surface elasticity causes higher axial stress values. 
Additionally, the surface elasticity reduces the sharp change in cross-section at the ends. 
Non-dimensionalized force-displacement data is shown in Fig.~\ref{fig:force-displacement-FEM} for three representative sets of surface parameters and three different element sizes ($l/R=0.04,\;0.02\text{ and }0.01$) to ensure the mesh convergence. 
Force-displacement behavior becomes non-monotonic for high surface tension values and is stabilized by surface shear modulus. 
Thin lines represent analytical results from \cite{zafar2021a} for ideal case (i.e., when radial slip is allowed at the ends). 
There is a significant difference between force responses of radially constrained cylinder and the ideal situation. 
Additionally, the non-monotonic force-displacement behavior is observed for $\overline{\sigma}_o=5,\overline{\mu}_s=0$ with radially constrained ends  (thick dotted green curve) in contrast to radially unconstrained ends (thin dotted green curve). 
It is noted that \cite{xuan2017} has reported monotonous behavior for $\overline{\sigma}_o<5.66$ when radial slip is allowed.

The data obtained from the FE simulations is used to train the ANN, as described in Section \ref{Sec:training}.

\section{Results}
\label{Sec:Results}
\subsection{Training of Artificial Neural Network}
\label{Sec:training}
\begin{figure}
    \begin{center}
    \includegraphics[width=0.5\linewidth]{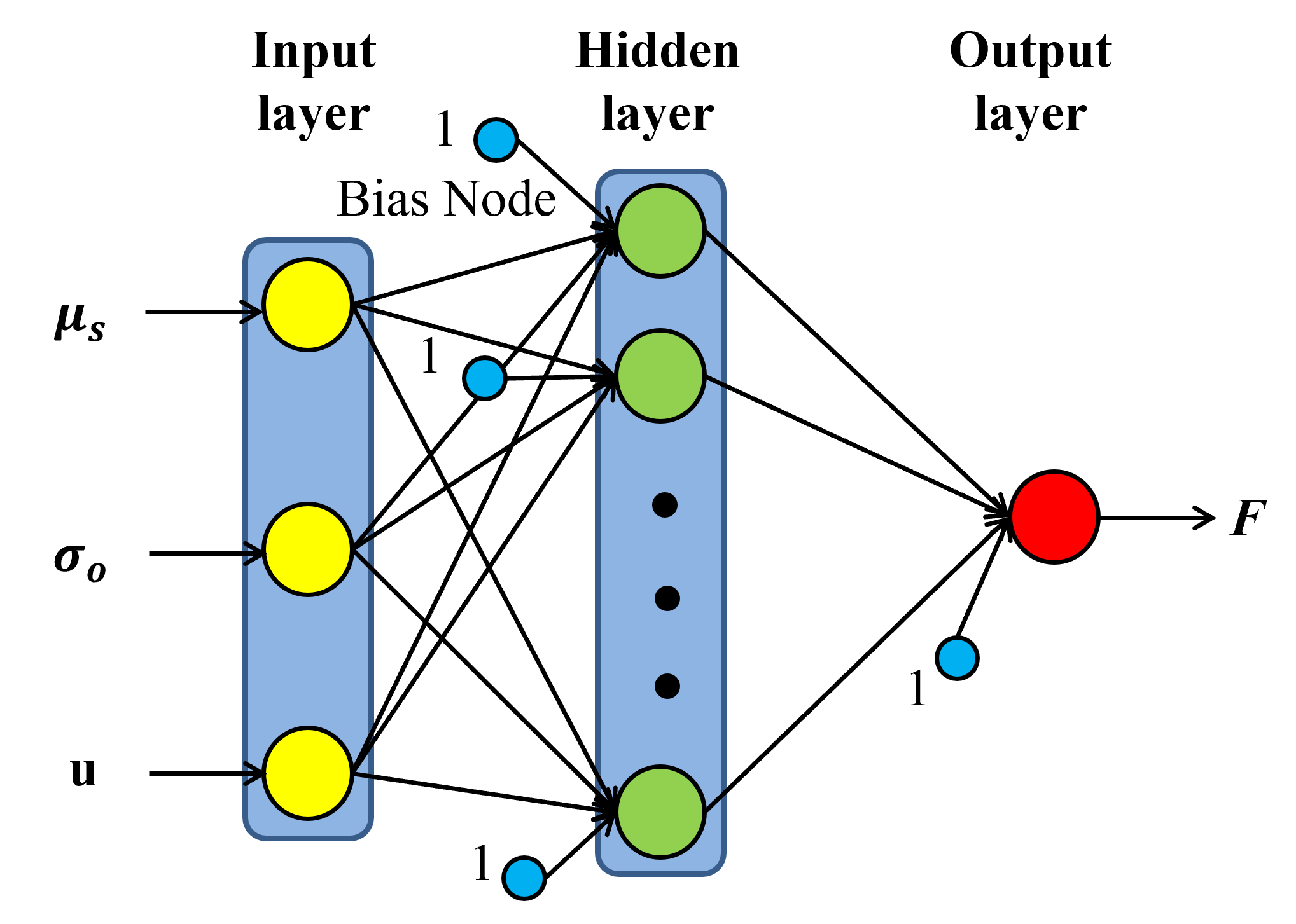}
    \caption{Schematic of ANN model.}
    \label{fig:ANNSchematic}
    \end{center}
\end{figure}

A feed-forward artificial neural network (ANN) with backpropagation is employed as a surrogate forward model for predicting the axial force response. The network architecture, illustrated in Fig.~\ref{fig:ANNSchematic}, is trained using the Levenberg–Marquardt optimization algorithm, chosen for its robustness and rapid convergence in nonlinear regression problems. The complete dataset is randomly partitioned into training (70\%), validation (15\%), and testing (15\%) subsets. A single hidden layer comprising 20 neurons with hyperbolic tangent sigmoid activation is adopted, while a linear activation function is used in the output layer. The number of neurons is selected based on preliminary trials to achieve a balance between prediction accuracy and computational efficiency. Termination criteria for training are: (i) a maximum of 1000 epochs, (ii) attainment of the prescribed performance goal (MSE = 0), (iii) an increase in validation error for six consecutive checks, or (iv) a performance gradient below 
$1\times 10^{-7}$.

The predictive capability of the trained ANN is evaluated using the mean absolute error (MAE), mean squared error (MSE), root mean squared error (RMSE), and the coefficient of determination ($R^2$). These performance metrics are summarized in Table~\ref{tab:Performance} for the training, validation, and testing datasets. To ensure robustness, the training procedure is repeated for several random initializations of network weights, yielding consistent performance in all cases. The ANN exhibits low MAE and RMSE values across all datasets relative to the maximum non-dimensional force values ($\sim700$), indicating accurate prediction of the force–displacement response. The small MSE values further confirm stable numerical convergence and effective learning.
\begin{table}[]
    \centering
    \caption{Performance indices of ANN Model}
    \begin{tabular}{c c c c c c}
        \hline
                 &  MAE     & MSE      & RMSE      & $R^2$ \\
        \hline
        Training  & 0.012464 & 0.000289 & 0.016998 & 1     \\
        Testing   & 0.014215 & 0.000361 & 0.019006 & 1     \\
        Validation& 0.013474 & 0.000335 & 0.018292 & 1     \\
        \hline
    \end{tabular}
    \label{tab:Performance}
\end{table}

\begin{figure}
    \begin{center}
    \subfigure[]{\includegraphics[width=0.45\linewidth]{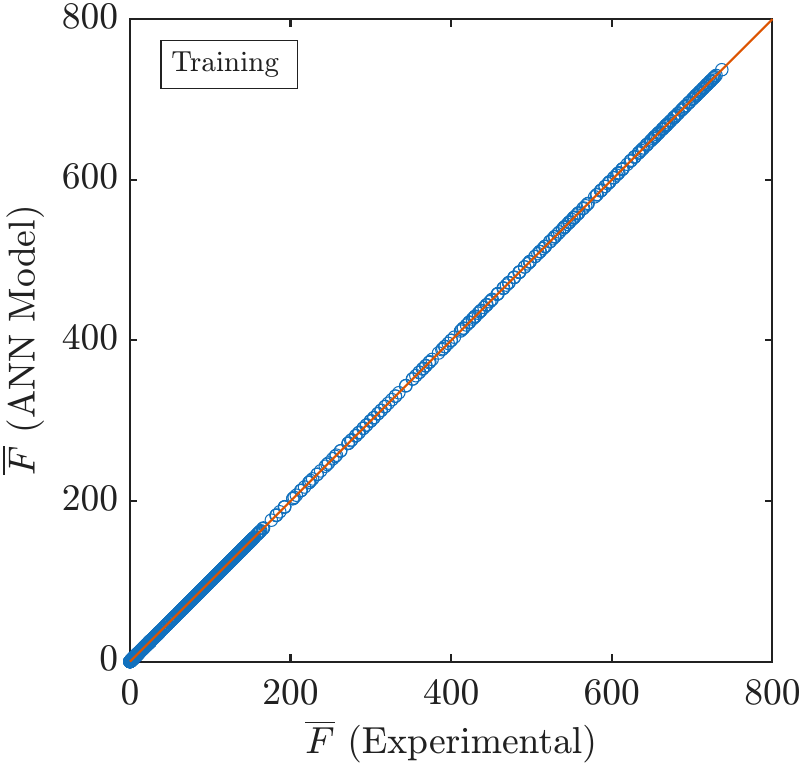}}
    \subfigure[]{\includegraphics[width=0.45\linewidth]{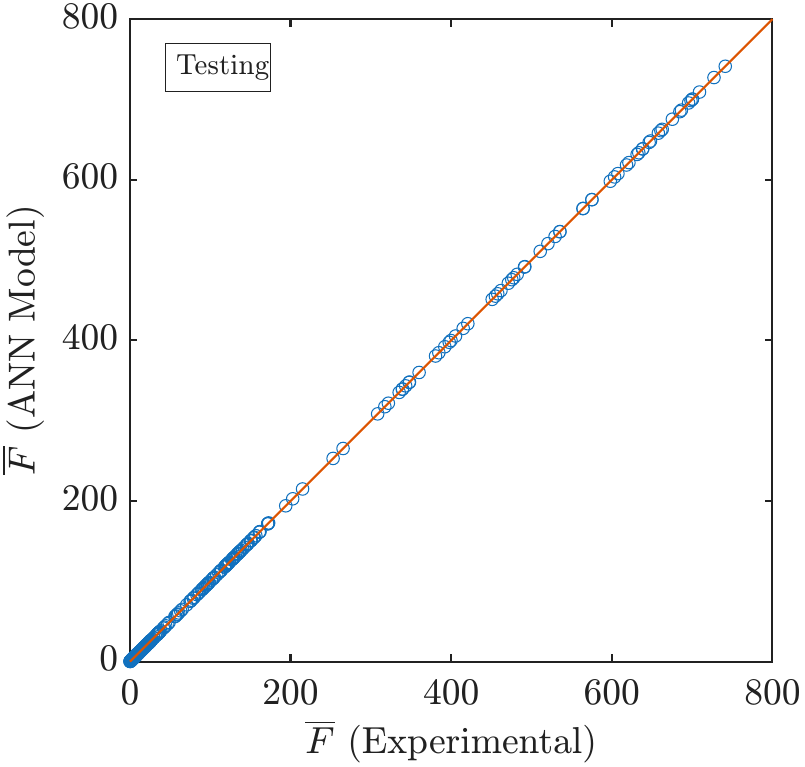}}
    \subfigure[]{\includegraphics[width=0.45\linewidth]{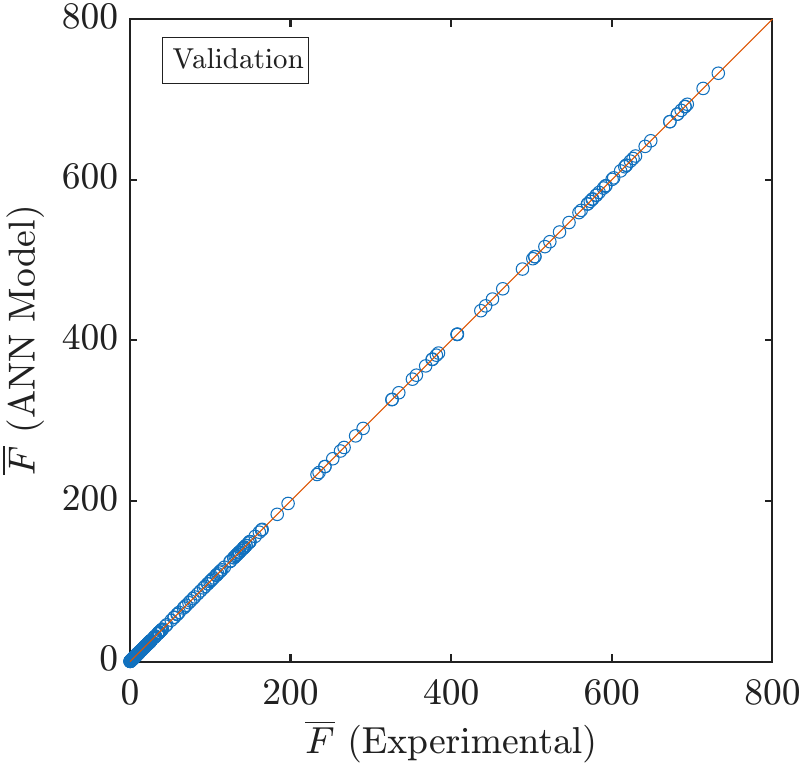}}
    \caption{Non-dimensionalized force obtained from ANN model for given displacement and material parameters against the corresponding experimental (or simulated) force for (a) Training, (b) Testing and (c) Validation data. Solid line shows the $45^\circ$ reference line for perfect agreement.}
    \label{fig:RegressionPlots}    
    \end{center}
\end{figure}

The coefficient of determination ($R^2$) achieved a value of unity for the training, testing, and validation datasets, highlighting an almost perfect correlation between ANN-predicted and FE-generated force values. 
This strong agreement is further illustrated by the regression plots shown in Fig.~\ref{fig:RegressionPlots}, where the predicted force values closely follow the ideal $45^\circ$ reference line for all three datasets. The close clustering of data points around the reference line indicates that the trained ANN effectively captures the nonlinear dependence of axial force on displacement and surface parameters.

\begin{figure}
    \begin{center}
    \subfigure[]{\includegraphics[width=0.45\linewidth]{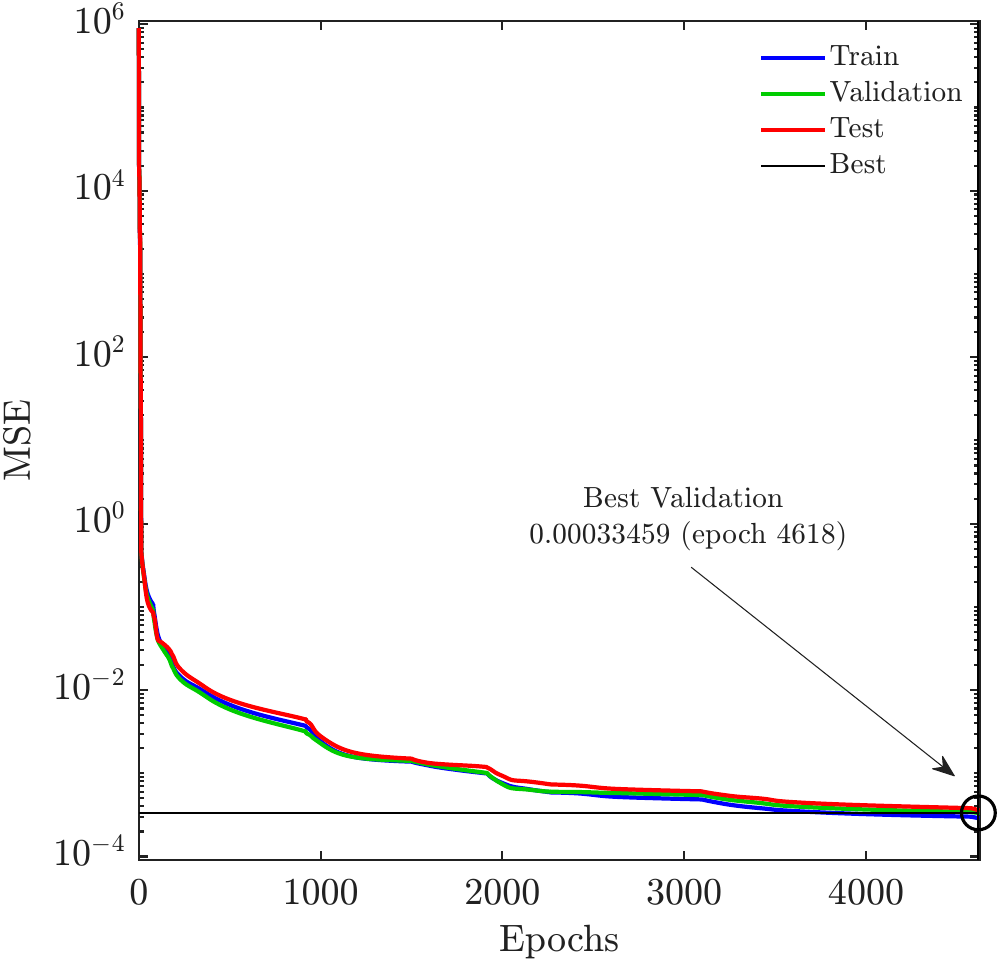}}
    \subfigure[]{\includegraphics[width=0.45\linewidth]{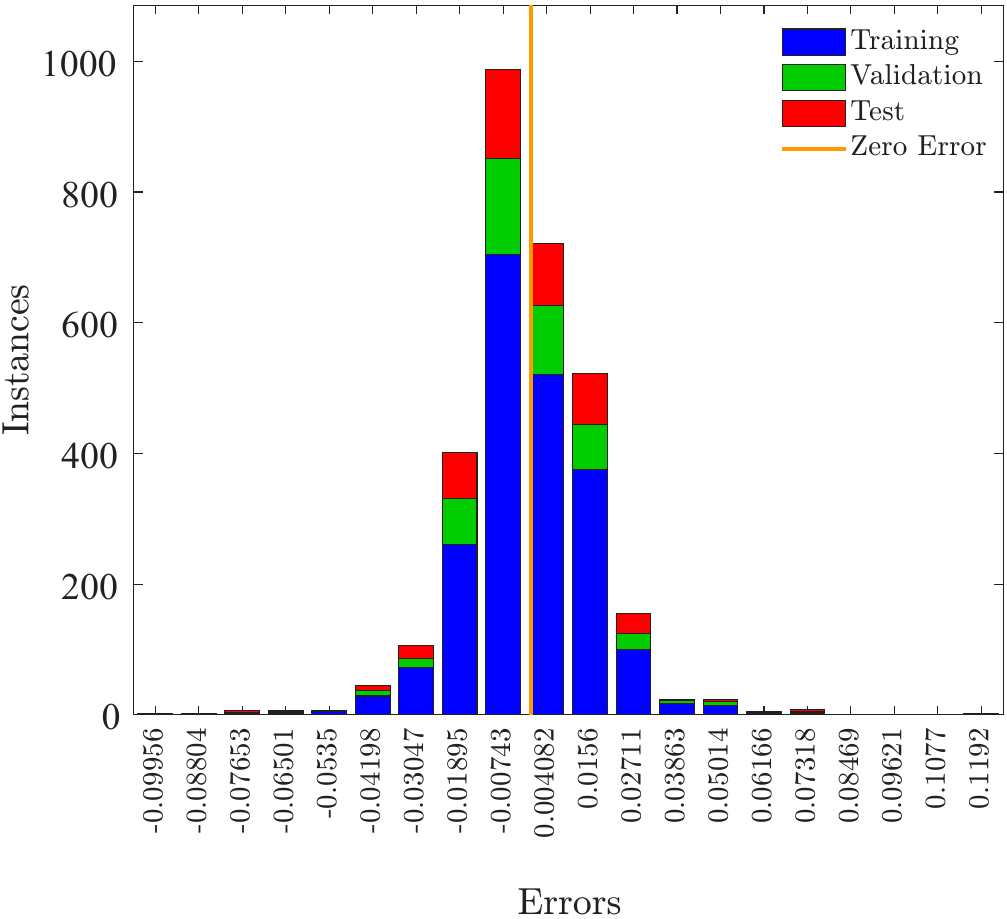}}
    \caption{(a) Convergence plot of ANN model and (b) Error Histogram for training, testing and validation data.}
    \label{fig:ConvergencePlot}    
    \end{center}
\end{figure}

In addition, the convergence behavior shown in Fig.~\ref{fig:ConvergencePlot}(a) exhibits smooth learning characteristics, with comparable trends observed for the training, validation and testing losses, indicating effective mitigation of overfitting. The corresponding error distribution in Fig.~\ref{fig:ConvergencePlot}(b) is nearly symmetric and centered around zero, suggesting the absence of systematic prediction bias. Overall, the results demonstrate that the developed ANN model provides an accurate and reliable representation of the tensile response of soft cylinders with elastic surfaces, with good generalization across the considered range of surface parameters and deformation levels.

\subsection{Identification of Surface Parameters}
A GA–based surrogate optimization framework is developed to inversely identify surface material parameters that can significantly influence the deformation behavior of soft solids (see, Fig.~\ref{fig:force-displacement-FEM}). 
The inverse problem aims to determine optimal values of non-dimensionalized surface shear modulus, $\overline{\mu}_s$, and surface tension, $\overline{\sigma}_o$, using target force-displacement data. Owing to the nonlinear coupling between bulk and surface effects, the resulting inverse problem is non-convex, therefore, the GA is chosen as it ensures efficient and reliable exploration of the parameter space. 

Initially, synthetic data from FE simulations are used to asses the performance of the framework under controlled condition. To mimic experimental uncertainty, subsequent results consider the same data with added noise, illustrating the robustness of the proposed approach for potential application to experimental measurements. 

\begin{figure}
    \begin{center}
    \includegraphics[width=0.65\linewidth]{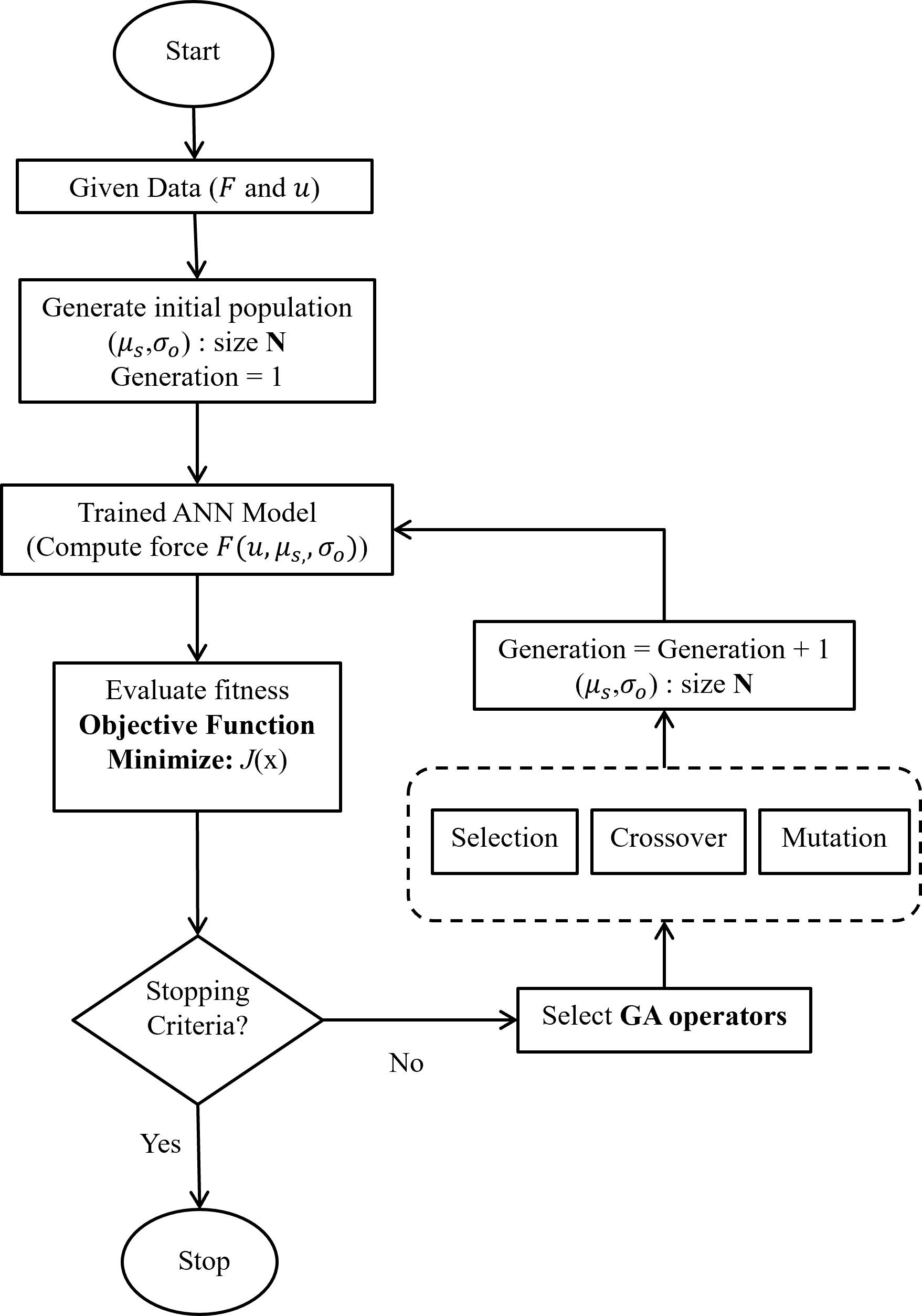}
    \caption{Flow chart of the GA–ANN optimization framework for inverse identification of surface elastic parameters}
    \label{fig:flowchart}    
    \end{center}
\end{figure}

To achieve computational efficiency, a previously trained ANN model is employed as a surrogate forward model within the GA optimization loop. This surrogate replaces repeated evaluations of the forward mechanical model, enabling rapid fitness evaluation during the GA search process. The overall optimization workflow is illustrated in the flowchart shown in Fig.~\ref{fig:flowchart}.

\subsubsection{Objective Function Formulation}
The optimization problem is posed as a minimization of a scalar fitness function. Accordingly, the inverse identification problem is formulated by defining the fitness function as the mean squared error (MSE) between the target and ANN-predicted forces:
\begin{eqnarray}
    &&\text{Minimize } J(\mu_s,\sigma_o),\nonumber\\
    &&J(\mu_s,\sigma_o)=\dfrac{1}{n}\sum\limits_{i=1}^n\left(F_i^{\text{tar}}-F_i^{\text{ANN}}\right)^2,
    \label{eq:objective}
\end{eqnarray}
where, $F_i^{\text{tar}}$ denotes the target force corresponding to the $i^{\text{th}}$ displacement, $F_i^{\text{ANN}}$ is the force predicted by the ANN surrogate model for the same displacement and $n$ is the total number of data points.

\subsubsection{Genetic Representation and Constraints}
Each candidate solution in the GA population is represented by a real-valued decision vector:
\begin{eqnarray}
    \bm{\mathrm{x}}=[\overline{\mu}_s\;\; \overline{\sigma}_o]
\end{eqnarray}
subject to the bound constraints
\begin{eqnarray}
    \mu_{\text{min}}<\overline{\mu}_s<\mu_{\text{max}}\nonumber\\
    \sigma_{\text{min}}<\overline{\sigma}_o<\sigma_{\text{max}}
\end{eqnarray}
These bounds were prescribed based on physical admissibility and prior knowledge of the surface material properties. In the present study, we have selected the values as $\mu_{min}=0$, $\mu_{max}=50$, $\sigma_{min}=0$ and $\sigma_{max}=50$.

\subsubsection{GA operators}
The genetic operators employed in the GA optimization framework are:

\noindent{\bf Selection:} Parent selection is performed using stochastic uniform selection, in which individuals with lower fitness values are assigned higher selection likelihoods while maintaining uniform sampling across the population. This selection strategy promotes population diversity and reduces the risk of premature convergence.

\noindent{\bf Crossover:} Crossover is applied to selected parent solutions using a real-coded crossover scheme. Given two parent vectors $\bm{\mathrm{x}}^{p_1}$ and $\bm{\mathrm{x}}^{p_2}$, the resulting offspring vector $\bm{\mathrm{x}}^{c}$ is expressed as:
\begin{eqnarray}
    \bm{\mathrm{x}}^c=\alpha \bm{\mathrm{x}}^{p_1} + (1-\alpha) \bm{\mathrm{x}}^{p_2};\;\;\; \alpha\in[0,1],
\end{eqnarray}
where $\alpha$ is a random mixing coefficient controlling the genetic contribution of each parent.

\noindent{\bf Mutation:} Mutation is introduced to preserve genetic diversity by applying small random perturbations to the offspring:
\begin{eqnarray}
    \bm{\mathrm{x}}^m=\bm{\mathrm{x}}^c+\bm{\epsilon},
\end{eqnarray}
where, $\bm{\epsilon}$ is a random perturbation vector with zero mean and bounded magnitude, ensuring feasibility of the mutated solution.

\subsubsection{Convergence and Stability of Optimization Framework}
To assess the effectiveness and stability of the proposed inverse identification framework, the GA optimization is performed to identify the surface parameters for FE-generated force–displacement data using MATLAB’s Global Optimization Toolbox. A population size of 50 individuals is employed, while all other GA parameters, including crossover fraction, mutation operator, elitism, and stopping criteria, were retained at their default MATLAB settings. 
The objective function (Eq.~\ref{eq:objective}) is evaluated using force–displacement data sampled at 101 discrete displacement points. The GA optimization is terminated based on MATLAB’s default stopping criteria, namely a maximum of $(100 \times \text{\texttt{No.\ of Variables}})$ generations or stagnation of the best fitness value for 50 consecutive generations with a function tolerance of $10^{-6}$.

\begin{figure}
    \begin{center}
    \subfigure[]{\includegraphics[width=0.45\linewidth]{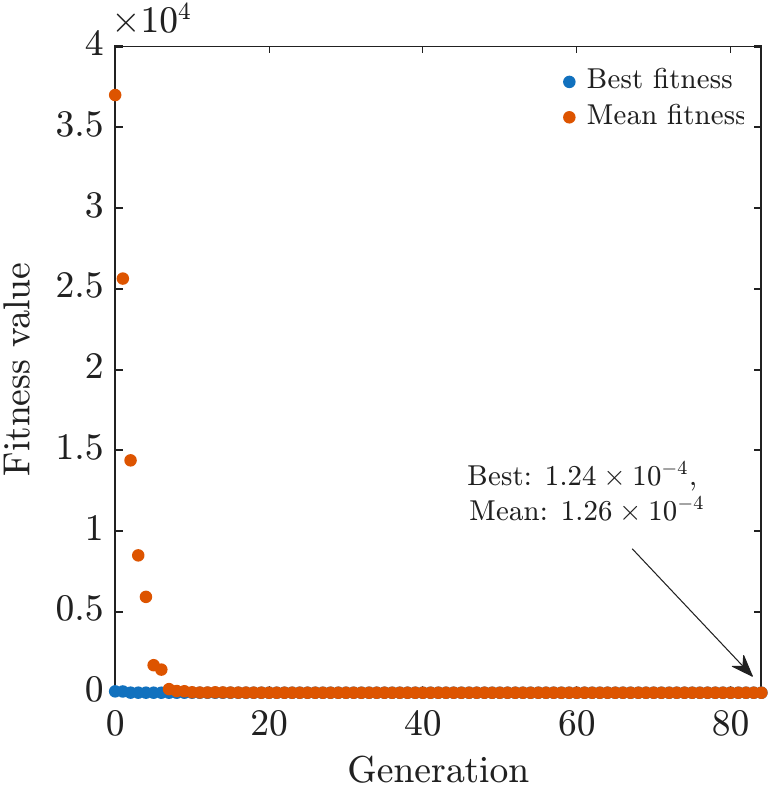}}
    \subfigure[]{\includegraphics[width=0.45\linewidth]{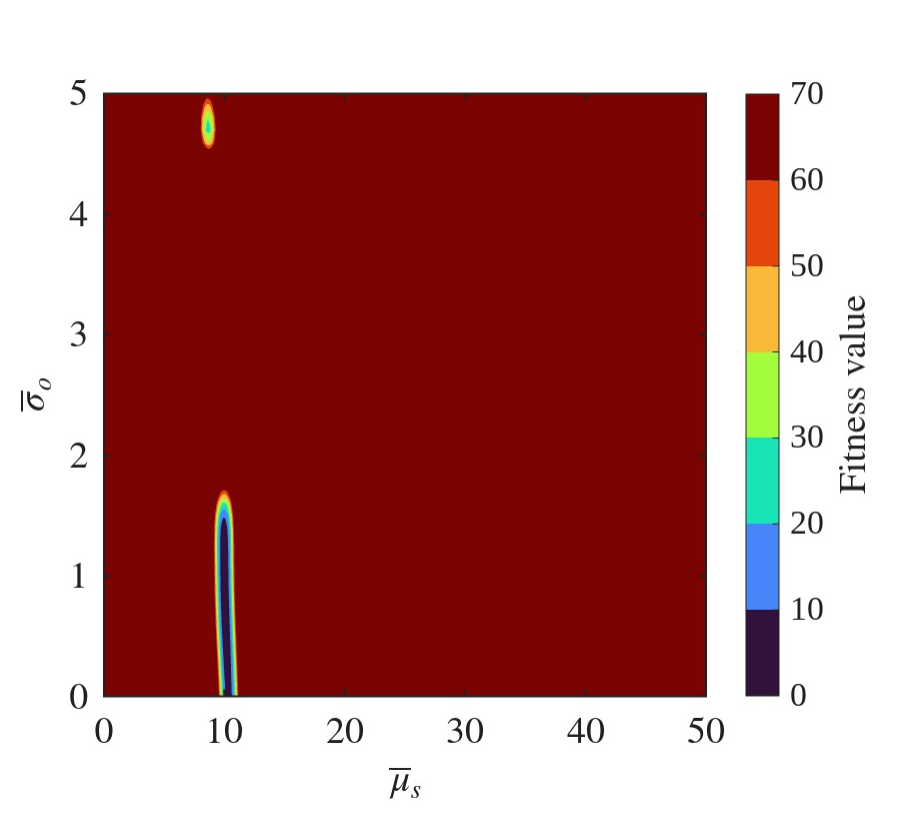}}
    \caption{(a) Evolution of best and mean fitness values with generations for $\overline{\mu}_s=10$, $\overline{\sigma}_o=5$ and (b) The Fitness value for $\overline{\mu}_s=10$, $\overline{\sigma}_o=1$ over the parameter plane.}
    \label{fig:evolution}
    \end{center}
\end{figure}

Figure~\ref{fig:evolution} (a) presents a representative convergence history, illustrating the evolution of the best and mean fitness values with generation number. Similar convergence behavior is observed for other parameter combinations considered in this study. A rapid reduction in the objective function during the early generations indicates effective global exploration of the parameter space, followed by gradual stabilization as the algorithm converges toward optimal solutions. The close agreement between the best and mean fitness curves at later generations reflects stable convergence and population consistency. The optimized values of $\overline{\mu}_s$ and $\overline{\sigma}_o$ corresponding to the best individual are summarized in Table \ref{tab:mu_sigma_comparison}, showing close agreement with the actual parameter values used to generate the target data. These results demonstrate that the proposed GA–ANN framework can reliably identify surface elastic parameters from noiseless force–displacement data within the prescribed bounds. Additionally, the fitness value over parameter space for a particular values of surface parameters shows two local minima in Fig.~\ref{fig:evolution} (b), highlighting the non-convex nature of the inverse problem and justifying the use of a GA-based optimization strategy.
\begin{table}[h]
\centering
\caption{Actual and predicted values of surface parameters with numerically generated force-displacement data.}
\label{tab:mu_sigma_comparison}
\renewcommand{\arraystretch}{1.2}
\begin{tabular}{c|c c c|c c c}
\hline
Sl. No. & $\overline{\mu}_s$ (Actual) & $\overline{\mu}_s$ (Predicted) & \% Error & $\overline{\sigma}_o$ (Actual) & $\overline{\sigma}_o$ (Predicted) & \% Error \\
\hline
1  & 0   & 0       & -   & 0   & 0.0000 & -    \\
2  & 0   & 0       & -   & 0.1 & 0.0989 & 1.10  \\
3  & 0   & 0.0005  & -   & 1   & 0.9985 & 0.15  \\
4  & 0.1 & 0.1010  & 1.00 & 0.1 & 0.1010 & 1.00  \\
5  & 0.1 & 0.1025  & 2.50 & 5   & 4.9997 & 0.01  \\
6  & 1   & 1.0003  & 0.03 & 0   & 0.0001 & -    \\
7  & 1   & 1.0005  & 0.05 & 0.1 & 0.0961 & 3.90  \\
8  & 1   & 0.9973  & 0.27 & 1   & 1.0041 & 0.41  \\
9  & 10  & 9.9983  & 0.02 & 0.1 & 0.1020 & 2.00  \\
10 & 10  & 10.0004 & 0.00 & 1   & 0.9978 & 0.22  \\
11 & 10  & 9.9996  & 0.00 & 5   & 4.9998 & 0.00  \\
12 & 50  & 49.9999 & 0.00 & 0.01& 0.0103 & 3.00  \\
13 & 50  & 49.9994 & 0.00 & 1   & 1.0024 & 0.24  \\
14 & 50  & 50.0000 & 0.00 & 5   & 5.0000 & 0.00  \\
\hline
\end{tabular}
\end{table}

In addition to convergence behavior, computational efficiency is a key motivation for employing a surrogate-assisted optimization strategy. In the present GA-based inverse identification, each optimization run typically converges within approximately 80 generations with a population size of 50, resulting in about 4000 forward model evaluations per run. A single nonlinear FE-based forward simulation, which computes force responses at 100 discrete displacement points, requires approximately 40 s of wall-clock time on a standard workstation. Consequently, a direct FE-in-the-loop GA optimization would require roughly $1.6\times10^5$ s ($\approx44$ h) per optimization run. In contrast, evaluation of the trained ANN surrogate requires approximately 0.02 s for the same input-output mapping, leading to a total wall-clock time of about 80 s ($\approx1.3$ min) per optimization run. This corresponds to an effective speed-up of approximately three orders of magnitude ($\sim2000\times$) compared to FE-based optimization. 

\subsubsection{Convergence Behavior for simulated data with noise}
In order to assess the robustness of the proposed inverse method for experimental data which includes noise, artificial noise is introduced into the FE-generated force data. Specifically, a uniformly distributed random error of $\pm5\%$ is added to the force values as
\begin{eqnarray}
    F^*(u)=F(u)+F(u)\;\mathcal{U}(-0.05,0.05),
\end{eqnarray}
simulating measurement noise commonly encountered in experiments. Here, $\mathcal{U}(-0.05,0.05)$ is uniformly distributed random error in the range $-0.05$ to $0.05$. The optimization procedure is then repeated using these perturbed datasets.

The predicted surface parameters obtained from the noisy dataset are summarized in Table \ref{tab:mu_sigma_erronious}. Despite the presence of noise, the framework is able to recover the surface shear modulus and surface tension with reasonable accuracy across a wide range of parameter combinations. While an increase in prediction error is observed compared to the noiseless case particularly for surface tension, the overall trends indicate stable convergence and satisfactory parameter identification. These results highlight the robustness of the proposed GA–ANN approach and suggest its potential applicability to experimental data where noise and uncertainty are unavoidable.
\begin{table}[h]
\centering
\caption{Actual and predicted values of surface parameters for erroneous  force-displacement data.}
\label{tab:mu_sigma_erronious}
\renewcommand{\arraystretch}{1.2}
\begin{tabular}{c|c c c|c c c}
\hline
Sl. No. & $\overline{\mu}_s$ (Actual) & $\overline{\mu}_s$ (Predicted) & \% Error & $\overline{\sigma}_o$ (Actual) & $\overline{\sigma}_o$ (Predicted) & \% Error \\
\hline
1  & 0   & 0.0000  & -   & 0   & 0.0000 & -     \\
2  & 0   & 0.0056  & -   & 0.1 & 0.0938 & 6.16   \\
3  & 0   & 0.0060  & -   & 1   & 0.9841 & 1.59   \\
4  & 0.1 & 0.1003  & 0.30 & 0.1 & 0.1037 & 3.72   \\
5  & 0.1 & 0.0990  & 1.01 & 5   & 5.0000 & 0.00   \\
6  & 0.1 & 0.1010  & 1.01 & 1   & 0.9843 & 1.57   \\
7  & 1   & 0.9992  & 0.08 & 0   & 0.0011 & -     \\
8  & 1   & 0.9988  & 0.12 & 0.1 & 0.1055 & 5.49   \\
9  & 1   & 0.9822  & 1.78 & 1   & 1.0291 & 2.91   \\
10 & 10  & 10.0652 & 0.65 & 0.1 & 0.0950 & 5.00   \\
11 & 10  & 9.9879  & 0.12 & 1   & 1.0583 & 5.83   \\
12 & 10  & 9.9085  & 0.92 & 5   & 5.0000 & 0.00   \\
13 & 50  & 49.8534 & 0.29 & 1   & 0.9242 & 7.58   \\
14 & 50  & 50.0000 & 0.00 & 5   & 4.8967 & 2.07   \\
\hline
\end{tabular}
\end{table}

\subsubsection{Repeatability and Uncertainty Analysis of Identified Surface Parameters}
In order to quantify the statistical uncertainty associated with the inverse identification process in the presence of stochastic noise, the GA–ANN optimization framework is executed independently $n=30$ times, each with a newly generated noisy force–displacement dataset for the same set of surface parameters. This analysis enables assessment of the repeatability and robustness of the identified surface parameters under realistic measurement variability.

Let $\hat{y}_i$ denote the predicted parameter ($\overline{\mu}_s$ or $\overline{\sigma}_o$) from the $i^{\text{th}}$ independent optimization run. The mean predicted response is calculated as:
\begin{eqnarray}
    \overline{y}=\dfrac{1}{n}\sum\limits_{i=1}^{n}\hat{y}_i.\nonumber
\end{eqnarray}
The variability among predictions is quantified using the sample standard deviation as:
\begin{eqnarray}
    s=\sqrt{\dfrac{1}{n-1}\sum\limits_{i=1}^{n}\left(\hat{y}_i-\overline{y}\right)^2}.\nonumber
\end{eqnarray}
Since the true variance of the predicted response is unknown, a t-distribution with $(n-1)$ degrees of freedom is employed to construct a confidence interval ($CI$) for the mean predicted response:
\begin{eqnarray}
    CI_{1-\alpha}=\overline{y}\pm \dfrac{s}{\sqrt{n}}t_{\left(\alpha/2,n-1\right)},\nonumber
\end{eqnarray}
where, $t_{\left(\alpha/2,n-1\right)}$ denotes the critical value of the Student’s t-distribution corresponding to a two-tailed confidence level $1-\alpha$ with $n-1$ degrees of freedom.
This confidence interval quantifies the uncertainty associated with the mean estimated parameter value and serves as a measure of the statistical robustness of the proposed optimization framework under stochastic noise. 

The mean values, standard deviations, and corresponding $95\%$ confidence interval based on the Student’s t-distribution for $\overline{\mu}_s$ and $\overline{\sigma}_o$, obtained from 30 independent runs, are reported in Table \ref{tab:CI}. The confidence intervals for both the parameters are narrow and symmetrically distributed around the mean estimates, indicating low variability and strong repeatability of the inverse identification process. The close agreement between the mean predicted values and the actual parameters further confirms the statistical stability and reliability of the GA–ANN framework in the presence of stochastic noise.
\begin{table}[h]
\centering
\caption{Confidence Interval for prediction of surface parameters.}
\label{tab:CI}
\renewcommand{\arraystretch}{1.2}
\begin{tabular}{c c c c c c}
\hline
Parameter & Actual Value & Mean & Standard Deviation & Lower Bound & Upper Bound \\
\hline
$\overline{\mu}_s$ & 0.1  & 0.1007  & 0.0048 & 0.0989 & 0.1025 \\
$\overline{\sigma}_o$ & 1  & 0.9992 & 0.0119 & 0.9948 & 1.0037 \\
\hline
\end{tabular}
\end{table}

\section{Conclusion}
\label{Sec:conclusion}
An ANN model has been successfully trained and validated using force–displacement data generated from nonlinear FE simulations over a wide range of surface elastic parameters. 
The trained ANN exhibited excellent predictive accuracy without signs of overfitting or systematic bias. A GA–based surrogate optimization framework is subsequently developed by embedding the trained ANN as a forward model to inversely identify surface parameters from force–displacement data. The framework is tested using both noiseless and noisy datasets corresponding to known surface parameter values.

The results demonstrate that the proposed optimization framework is robust, efficient, and statistically consistent for the inverse identification of surface elastic parameters in soft solids. Replacing the nonlinear FE model with the ANN surrogate within the optimization loop significantly reduces the computational cost while maintaining high accuracy, making the approach well suited for repeated forward evaluations required in inverse analyses.

The proposed methodology can be directly applied to experimental force–displacement data for determining surface elastic parameters of soft materials. Furthermore, the framework can be readily extended to incorporate more complex bulk \cite[see, e.g.,][]{Hossain2013,ANSSARI2023} and surface constitutive models. It also holds potential for application to other calibration experiments in soft matter mechanics, such as needle-induced cavitation and other deformation-based characterization techniques.

\section{Acknowledgment}
MRZ acknowledges the Anusandhan National Research Foundation (ANRF), Government of India, for financial support through the Core Research Grant (CRG/2023/000080), which made this work possible.

\bibliography{ref}
\end{document}